\documentstyle[11pt,paspconf,psfig]{article}

\begin{document}

\title{The Interaction of the Disk with the Halo}
\author{Mordecai-Mark Mac Low\altaffilmark{1}}
\affil{MPI f\"ur Astronomie, K\"onigstuhl 17, D-69117, Heidelberg, Germany}

\altaffiltext{1}{Address after 1 May 1999: Department of Astrophysics,
American Museum of Natural History, Central Park West at 79th Street,
New York, New York, 10024-5192, United States}

\begin{abstract}
In this review I focus on how massive stars in the disks of galaxies
determine the properties of the ionized, gaseous haloes of
star-forming galaxies.  I first examine the ionization state of the
halo, and conclude that diluted radiation from massive stars, along
with a possible secondary heating source such as shocks, appears
sufficient to explain the observations.  I then look at how single
superbubbles evolve, as they must provide one of the main mechanisms
for transfer of mass and energy to the halo, as well as opening up
lines of sight for ionizing radiation to escape from the disk.
Finally, I look at the interactions of superbubbles, as they will
rarely evolve in isolation, and describe recent hydrodynamical and MHD
simulations of the interstellar medium shaped by multiple superbubbles.
\end{abstract}


\section{Introduction}

The high temperatures and ionization fractions of the gaseous haloes
of star-forming galaxies must be explained by energetic processes,
most likely occurring in the disks of the galaxies (though external
sources could contribute).  The most likely energy source for these
processes are massive OB stars.  

Stellar winds and supernovae drive shock waves into the interstellar
medium (ISM) in the disk that can create superbubbles large enough to
blow out of the disk into the halo.  Numerical models of these
blowouts reveal that rather little mass is lifted out of the disk by
superbubbles (typically only a few percent of the total mass swept up
by the shell), although that may still be enough mass to dominate the
low-density halo.  However, radiation from massive stars that would
normally be absorbed by dust and gas in the galactic disk can escape
out into the halo through the holes carved by superbubbles.  This
radiation will ionize the halo and could even drive gas motions.  The
superbubble chimneys might also provide an avenue out of the disk for
metals produced by the SNe of the massive stars in the plane, allowing
metal redistribution, or, particularly for starburst galaxies, even
metal ejection.

Understanding the overall structure of a halo controlled by massive
stars requires understanding not just the structure of isolated
superbubbles in a smoothly stratified background medium, but also the
results of their interactions.  Numerical models have started to
appear that address this, though the challenge of including all the
relevant physics (magnetic fields, radiation, heating and cooling,
cosmic rays) remains a daunting one.  However, such models are
required to understand the actual structure of the chimneys that allow
the existence of the halo.  They will also tell us a great deal about
the structure of the ISM in the disk, naturally.

I also briefly consider models of the galactic halo that rely
primarily on magnetic field generation by gas motions in the disk
followed by reconnection heating in the halo or even ejection of mass
along open field lines in a galactic wind.  In these models, the
energy would primarily be carried by the magnetic field, although the
source of energy for both field generation in the disk and for
transport of field into the halo may remain the massive stars in the
disk.

\section{Ionized Gas}

It has become clear through the efforts of Reynolds (1984, 1989) and
others that a large fraction of the ionized gas in the ISM is
contained not in traditional H~{\sc ii} regions but rather in diffuse
ionized gas with a scale height of a kiloparsec or more, as reviewed
by Dettmar (1992).  This gas has now been demonstrated by the
Wisconsin H-Alpha Mapper (WHAM; Reynolds et al. 1998) and other
surveys to be highly structured, and not a smooth, stratified medium.

Similar layers of H$\alpha$-emitting ionized gas are seen in external
galaxies.  They are most easily observed in edge-on spiral galaxies
such as NGC~891 (Rand, Kulkarni, \& Hester 1990, Dettmar 1990).  Not
all spiral galaxies show an observable extraplanar layer of ionized
gas (Pildis, Bregman, \& Schombert 1994; Rand 1996).  Rather, the
occurrence and intensity of such layers appears to be directly
correlated to the star-formation rate per unit area in the disk of the
galaxy (Rand 1996, 1998b).

The energy requirement for ionization of the diffuse ionized gas in
the Galaxy, both in the plane and above it, is 100\% of the kinetic
energy contributed by supernovae, or 15\% of the energy in ionizing
radiation from hot OB stars (Reynolds 1984, 1987, Dettmar 1992).  As it
would be rather difficult to imagine a fully efficient method for
transferring energy from supernovae to ionization of the diffuse gas,
radiation from hot stars appears at first the most likely suspect.

However, this raises the question of how to get so much of the
ionizing radiation out of classical H~{\sc ii} regions into the
diffuse gas hundreds or even thousands of parsecs away from the stars.
Furthermore, the ionization state of the diffuse ionized gas at high
altitudes is higher than in gas in H~{\sc ii} regions in our own
(Reynolds 1985a, b) and external galaxies (e.g., Dettmar \& Schulz
1992, Golla, Dettmar, \& Domg\"orgen 1996), as shown by observations
of values of the line ratios [S~{\sc ii}]/H$\alpha$ and [N~{\sc
ii}]/H$\alpha$ as much as twice as high as in H~{\sc ii} regions.

A number of explanations of this problem have been advanced.
Extragalactic ionizing sources will contribute to the ionization, but
can ionize only very low column densities of neutral hydrogen (Maloney
1993, Corbelli \& Salpeter 1993), so they cannot explain most of the
diffuse ionized gas in the plane of the galaxy.  Local heating by
shocks is suggested by the high [S~{\sc ii}]/H$\alpha$ and [N~{\sc
ii}]/H$\alpha$ line ratioes, but again runs into the question of the
energy source.  It may well be an important secondary source of
ionization to help explain the high line ratioes above the plane of
the galaxy (e.g. Martin 1997, Rand 1998a).  Similarly, turbulent mixing
layers in the walls of superbubbles might contribute to the heating
(Begelman \& Fabian 1990; Slavin, Shull, \& Begelman 1993).
Scattering of ionizing radiation by dust should harden the radiation
field with altitude (Ferrara et al. 1996), explaining the
increasing line ratioes.  However, Domg\"orgen \& Dettmar (1997)
observed similar increases in line ratioes with altitude in the
metal-poor galaxy NGC~2188, where dust should have played a smaller
part, suggesting that this explanation fails.

Local heating by reconnection of twisted magnetic field lines is
another possibility that has only been modeled in principle (Birk,
Lesch, \& Neukirch 1998), as I discuss in the final section.
Similarly, dissipation of MHD turbulence has been proposed by Minter
\& Spangler (1997).  A more exotic proposal by Sciama (1990, 1998) is
that the ionization is due to the decay of neutrinos with a mass of
$27.4 \pm 0.2$ eV, a value that may soon be directly comparable with
experimental results from Super-Kamiokande, which has already measured
a difference in masses between electron and muon neutrinos of only 0.07 eV
(Fukuda et al. 1998).

The most convincing explanation for the behavior of the ionization
with altitude still seems to me to be simple dilution of the ionizing
radiation field with distance from the stellar sources (Domg\"orgen \&
Mathis 1994).  Although it cannot explain all the details, it appears
to account for the general pattern of ionization.  Some secondary
sources of heating or ionization may also be at work, but they do not
dominate the energetics.  This model also predicts that about 4\% of
the ionizing radiation from OB stars actually escapes the Galaxy
entirely.

There are now quantitative measurements of the amount of ionizing
radiation escaping the Galactic disk.  High-velocity H~{\sc
i} clouds provide one screen upon which escaping radiation can be
observed.  They are located several kpc above the Galactic disk
(Wakker \& van Woerden 1997) outside of the observed diffuse ionized
gas.  Observations of them in H$\alpha$ by Tufte, Reynolds, \& Haffner
(1998) and Bland-Hawthorn et al. (1998) have detected weak emission
associated with them. Similarly, the Magellanic Stream provides
another screen, from which Weiner \& Williams (1996) have also
detected weak H$\alpha$ emission.  These observations have been
interpreted by Bland-Hawthorn \& Maloney (1998) as implying the
equivalent of an optical depth $\tau =3$ in a uniform galactic disk.
This is fully consistent with the prediction of Domg\"orgen \&
Mathis (1994).

These models still neglect the question of exactly how the radiation
reaches great distances from OB associations.  Presumably the
existence of superbubbles and other large-scale structures helps
radiation to travel long distances.  However, although the propagation
of radiation in a uniform stratified disk has been studied (Miller \& Cox
1993; Dove \& Shull 1994), as has the radiation from thermally ionized
mixing layers in the walls of blown-out superbubbles (Begelman \&
Fabian 1990; Slavin, Shull, \& Begelman 1992), the propagation of
radiation through a medium structured by superbubbles has not
(although a poster by Basu et al. at this conference took first steps
in that direction).  Certainly, the observed morphology of the diffuse
ionized gas does not strongly resemble the ionization cones computed by
Miller \& Cox (1993) or Dove \& Shull (1994).


\section{Single Superbubbles}

\begin{figure}
\psfig{file=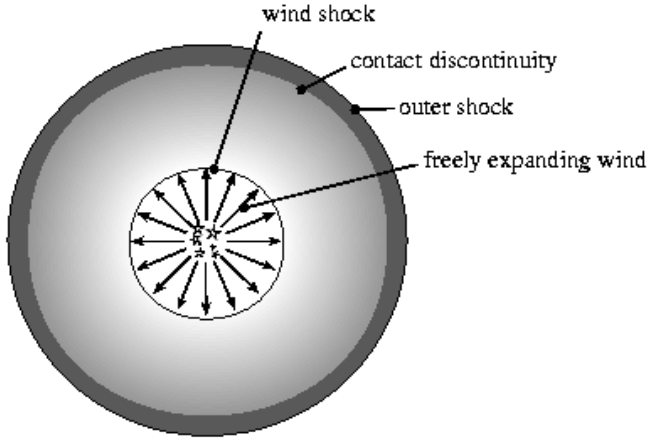,width=3.5in}
\caption[junk]{Structure of typical superbubble before blowout, showing freely
expanding winds and blast waves, inner termination shock, hot shocked
gas, contact discontinuity, cold, dense shell, and outer shock.}
\label{struct}
\end{figure}

Most ionizing OB stars remain in their parent associations, where ten
to a thousand massive stars have formed in a small region.  
All the stars within an OB association heavy enough to become SNe at
all will do so within 50 Myr of the formation of the association.  The
resulting superbubble must be at least part of the explanation of how
ionizing radiation can travel such long distances from the ionizing
stars.  Figure~\ref{struct} shows the structure of a typical
superbubble (Castor, McCray, \& Weaver 1975, McCray \& Kafatos 1987).
Winds and SN blast waves freely expand until they reach an inner
termination shock.  The kinetic energy of the expanding gas is
converted to thermal energy, producing a hot, shocked region.  The
pressure of this region drives a shock into the surrounding
interstellar gas, sweeping it up into a shell.  The temperature of the
hot, shocked region is held down to a few million degrees by
conductive evaporation off the swept-up shell.  In the rest of this
section, I describe the development of superbubbles as they interact
with stratified density distributions in dwarf and spiral galaxies.
(This topic has also recently been carefully reviewed by
Bisnovatyi-Kogan \& Silich 1995, with a focus on thin-shell methods).

The most basic question to ask is when will a superbubble blow out of
a stratified ISM---that is, when will it accelerate until it reaches a
uniform halo or intergalactic medium?  A dimensional analysis
calibrated by the semi-analytic model of Mac Low \& McCray (1988) can
give the answer (see Koo \& McKee 1990 for a similar treatment).  The
relevant scales for this problem are the mechanical luminosity of the
OB association $L_m$, the scale height of the density distribution
$H$, and the midplane density $\rho_0$, and pressure $P_0$.  The time
scale is then the time for a spherical superbubble to reach dimensions
comparable to a scale height (Castor {\em et al.} 1975), $t_1 \sim
H^{5/3} (\rho_0 / L_m)^{1/3}$.  The sound speed at the midplane $c_s =
(\gamma P_0 / \rho_0)^{1/2}$ gives a velocity scale.  This can be
compared to the velocity of the superbubble when it reaches a scale
height,$v_1 = (3/5) (H/t_1)$.  The Mach number of the superbubble when
it reaches a size comparable to a scale height $M_1 = v_1 / c_s$
determines whether a superbubble will ultimately blow out or be
pressure-confined by the external medium.

The actual calibration of this criterion must be done using
semi-analytic techniques.  Kompaneets (1960) showed that a blast wave
in a zero-pressure, exponential atmosphere will always blow out.  The
thin-shell approximation that he used can be adapted to the present
problem.  The assumptions used are that the shell of swept-up gas is
thin and that the interior pressure acting on it is uniform
everywhere.  The latter assumption is equivalent to assuming that the
sound speed in the interior is far greater than the expansion velocity
of the shell.  Mac Low \& McCray (1988) divided the shell up into a
number of segments and numerically integrated the equations of motion
of the shell in an arbitrary, finite-pressure, stratified atmosphere
to find its shape and expansion velocity.  They find that if a bubble
has $v_1/c_s > 5$ it will blow out of an exponential atmosphere.

When the shell begins to accelerate upward, it becomes Rayleigh-Taylor
unstable, as the rarefied interior effectively supports the dense
shell.  Numerical models show that the shell then fragments and
releases its internal pressure (Mac Low, McCray, \& Norman 1989, and
see the review by Tenorio-Tagle \& Bodenheimer 1988).  The subsequent
evolution has been analytically modeled by Koo \& McKee (1990).  It is
notable that when a shell blows out, only a very small fraction of its
mass is actually accelerated up at any significant velocity.  For
example, Mac Low {\em et al.} (1989) found that only 5\% of the shell
mass ended up in the portion of the shell accelerated upwards.

The models of superbubble blowout described so far assumed a
stratified but isothermal atmosphere, typically with a temperature of
$10^4$~K typical of the diffuse ionized gas.  The alternative extreme
was later studied, with a warm disk, and a hot halo with temperatures
of order $10^6$~K (Igumentschev, Shustov, \& Tutukov 1990;
Tenorio-Tagle, R\'o\.zyczka, \& Bodenheimer 1990), showing that
blowouts would eject small amounts of warm gas into a hot halo.  These
models also included simple approximations to thermal conduction,
although the conclusion was reached that this indeed doesn't make much
difference to shell structure.

Magnetic fields can strongly influence superbubble structure.  The
actual strength of the field in the Milky Way is estimated to be
between $3 \mu$G and $5 \mu$G.  If the field is towards the upper end
of this range, it is sufficiently strong to confine most superbubbles
and contain much of the kinetic energy injected by supernovae as
magnetic pressure ({\em e.g.} Slavin \& Cox 1992).  The behavior of
magnetized superbubbles was first studied using a semi-analytic
``thick-shell'' model by Ferri\`ere, Mac Low, \& Zweibel (1991), and
using a 2D numerical model by Tomisaka (1992).
Using slab-symmetric 2D models, Mineshige, Shibata, \&
Shapiro (1993) demonstrated that sufficiently strong magnetic fields
could prevent immediate blowout.  Tomisaka (1990, 1998) has now done
three-dimensional models demonstrating confinement, and also showing
that fields can lead to significant elongation of the bubbles, giving
them the structure of a flat slice of a mushroom (or, for readers who
have spent time in the US, a slice of Wonder Bread).

\begin{figure}[t]
\vspace*{-1in}
\psfig{figure=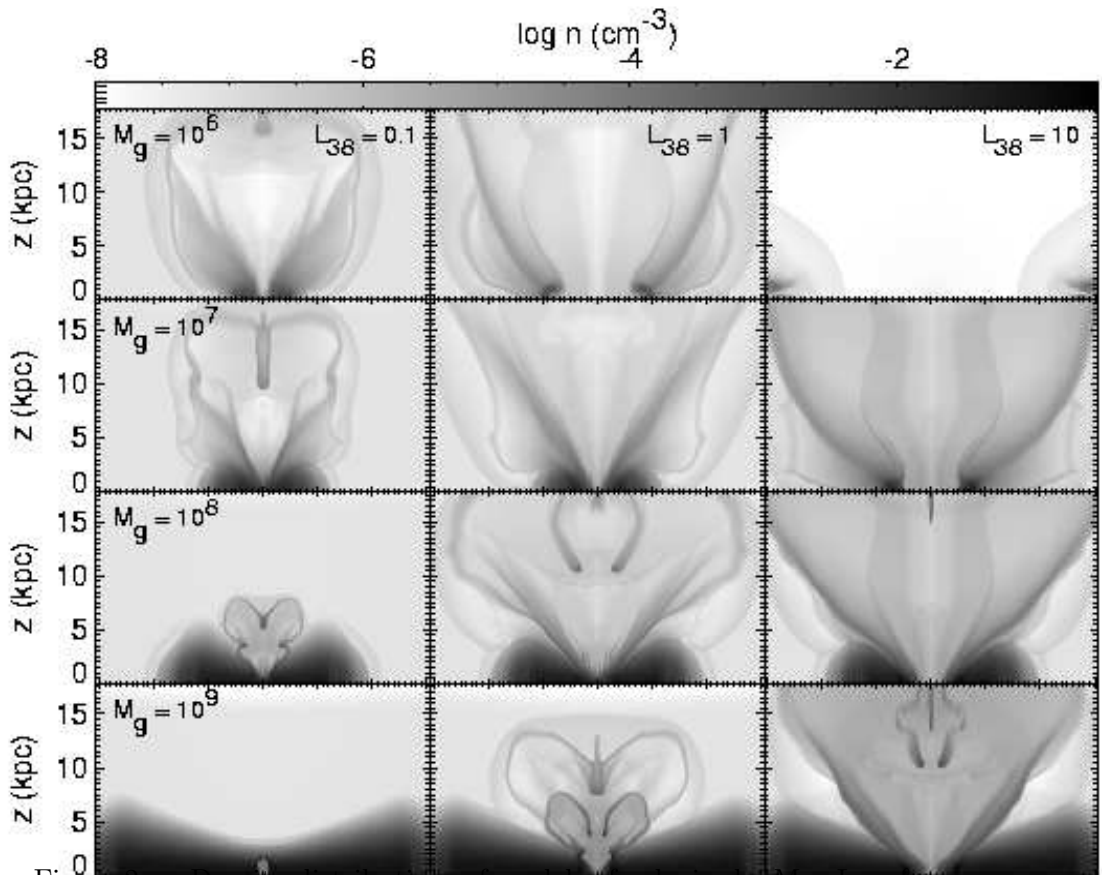,width=6in,rheight=4.4in}
\caption{Density distributions of models of galaxies by Mac Low \&
Ferrara (1999) with the given gas masses in solar masses, containing
starbursts of the given mechanical luminosities in units of $10^{38}$
erg~s$^{-1}$ at a time of 100 Myr, some 50 Myr after the end of energy
input from the starbursts, with the values of density given by the
colorbar at the top of the figure. The gravitational potentials from
dark matter haloes were included in these simulations.  Luminosities
are equivalent to supernovae every 30,000 to 3 million
yr.\label{starburst}}
\end{figure}

The most extreme case of the action of a single superbubble on the ISM
is the case of a starburst in a dwarf galaxy, where the combined
supernova explosions were even thought to be able to eject
the entire ISM from the potential well of the galaxy (Dekel \& Silk
1986).  This question has been studied using thin-shell models by
Silich \& Tenorio-Tagle (1998), and using two-dimensional gas
dynamical simulations by Mac Low \& Ferrara (1999).  Both groups show
that the coupling between the ISM in the galaxy and the energy
injected by the supernovae is poor because of the blowout of the
resulting superbubble from the galaxy.  Mac Low \& Ferrara (1999) show
that it is impossible with reasonable numbers of supernovae to
completely blow away the ISM for galaxies with gas mass $> 10^7
M_{\odot}$, as shown by the parameter study shown here in
Figure~\ref{starburst}.
On the other hand, they also show that metal-enriched ejecta from the
Type~II supernovae resulting from the starburst escape easily, as
first suggested by Vader (1986), providing a powerful selective
depletion mechanism that might explain the observed odd abundance
patterns.

\section{Many Superbubbles}

Superbubbles do not, in fact, expand into a smooth, stratified
interstellar medium, much as theorists might wish otherwise.  Rather,
they expand into a medium strongly structured by previous
superbubbles, isolated supernovae (especially in the halo; see Shelton
1998), ionization fronts, and radiative cooling.  Models must take
this into account in order to make contact with the observations at
anything more than the most qualitative level.

The first serious attempts to do this numerically were initiated by
Chiang \& Bregman (1988) using a multi-fluid code that followed the
gas and a star ``fluid'' in two dimensions.  The two fluids were
coupled by a density-dependent star formation term and by energy input
from the star fluid to the gas.  This approach was developed in a
series of papers by Rosen \& Bregman (1995; Rosen, Bregman, \& Norman
1993; Rosen, Bregman, \& Kelson 1996), adding physics, increasing
resolution, and computing observable quantities.  These computations
were the first to make clear that observed structures are not isolated
objects but just particularly recognizable parts of an active, complex
medium.  The difficulties encountered by radio observers in picking
structures out of the HI is not accidental, but due to the very
structure of the gas under observation.  Attempts to model these
structures as isolated objects will naturally then run into serious
difficulties and should be embarked on only with the greatest caution.

\begin{figure}[thbp]
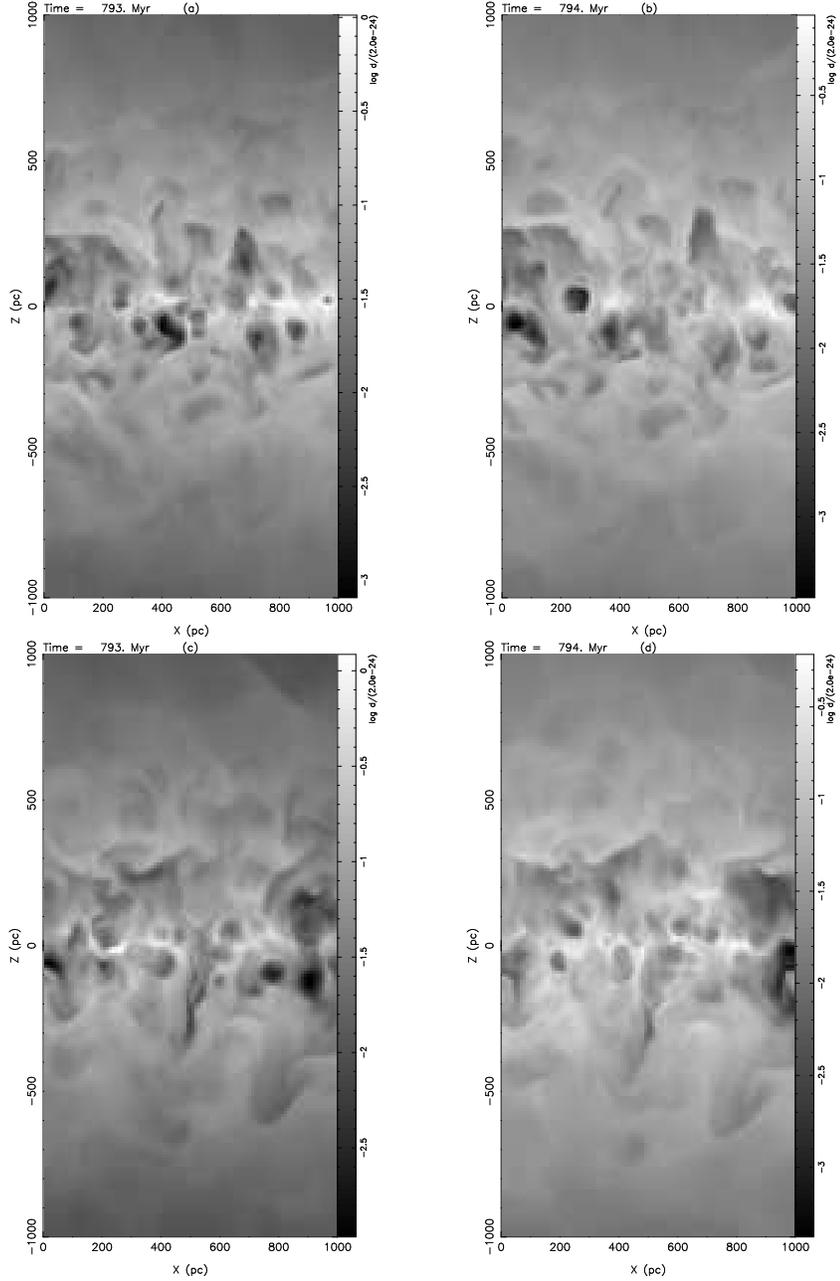

\centering{\hspace*{-0.1cm}}
\psfig{file=maclowm3.ps,angle=-90,width=11cm,clip=}
\psfig{file=maclowm4.ps,angle=-90,width=11cm,clip=}
\caption{\small Grey-scale maps showing the evolution of the disk gas
in three-dimensional models by Avillez, measured at two locations in
the disk: $y=480$ pc ((a) and (b)) and $y=980$ pc ((c) and (d)). The
times of these images are 793 Myr ((a) and (c)) and 794 Myr ((b) and
(d)). Globally the disk has similar structure in all the images,
although locally there are variations. Note the global expansion of
warm gas carving through the thick disk and the constant presence of the
wiggly cold disk in the midplane. Local outbursts such as the chimney
located at $x=500$ pc and $z=-100$ pc ((c) and (d)) are also
seen. \label{avillez}}
\end{figure}

Large-scale models are now being developed by several different
groups.  V\'azquez-Semadeni discusses his group's two-dimensional
models of the plane of the Galaxy including magnetic fields in these
proceedings.  The first three-dimensional models of the integrated
system were performed by Avillez, Berry, \& Kahn (1997) using an
adaptive mesh refinement technique to achieve 10 pc resolution.  They
explicitly included both correlated and random supernova energy input
in discrete events.  They found that a patchy, inhomogeneous Reynolds
layer shown in Figure~\ref{avillez} was naturally generated if enough
supernovae went off, quite regardless of the initial density profile
they started with.  Observed chimneys and worms formed due to
interactions of superbubbles, not as isolated superbubbles blowing out
of the disk, and so have rather different dynamics than would be
expected from isolated objects (Avillez 1998a, 1998b).

Magnetic fields will be important in determining the actual
interactions between superbubbles and other energetic structures in
the interstellar medium.  The models described up to this point have
neglected the effects of magnetic fields due to their numerical
difficulties, however.  Semi-analytic models of superbubbles driving a
galactic dynamo to maintain the magnetic field of the galaxy are
discussed by Ferri\`ere in these proceedings.  Her models, in
combination with observational constraints, yield the vertical
distribution of hot gas above the plane of the galaxy, with high
filling factors of hot gas occurring at altitudes of more than a scale
height or so (Ferri\`ere 1995, 1998).  This suggests that the observed
warm ionized gas at high altitudes might actually be confined to
thin sheets with very low filling factor.

\begin{figure}
\begin{center}
\hspace*{2cm}\psfig{file=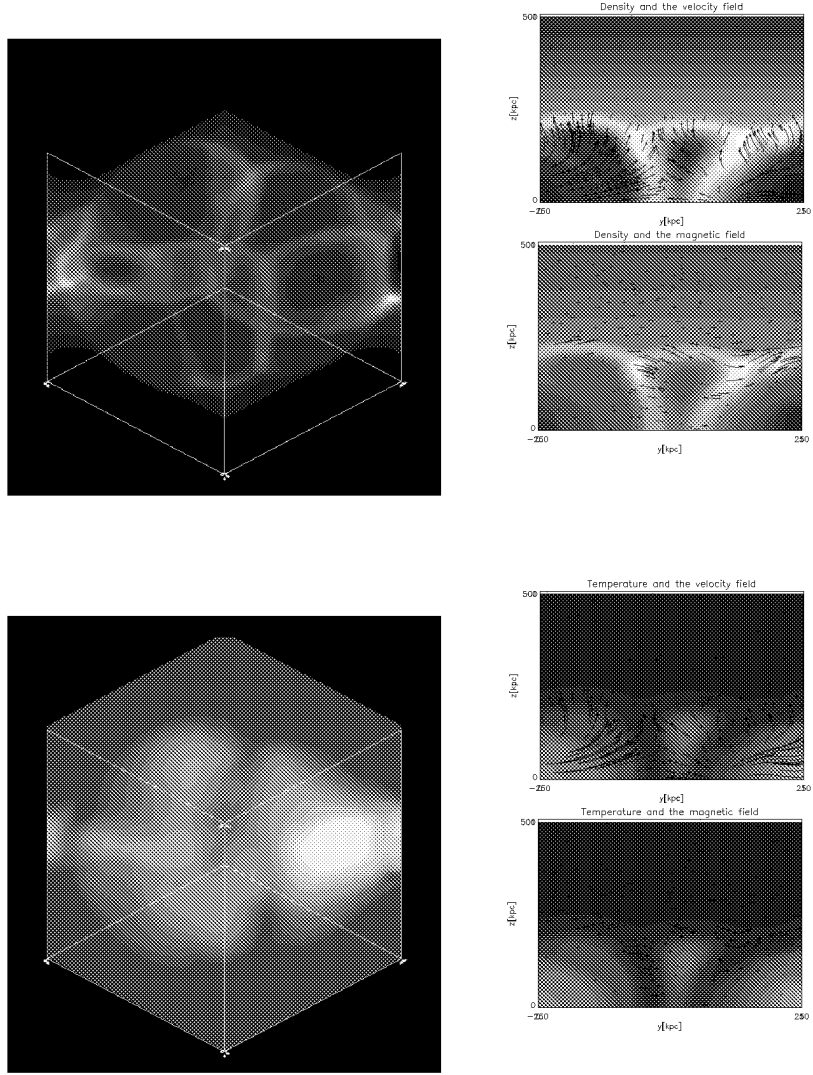}
\end{center}
\caption{A local 3D MHD simulation of the Galaxy by Korpi, {\em et
al.} (1998, also see this volume), with $L_x = L_y = \frac12 L_z = 0.5
kpc$.  The disk is heated by supernova explosions, which produce hot
bubbles surrounded by thin, dense and cool shells (see upper and lower
left panels: 3D plots of density and temperature are shown).
Supernovae can interact and form superbubbles: in the upper left
panel, smaller bubbles are produced by single supernovae, while the
larger ones contain multiple supernovae.  Due to density
stratification in the vertical direction, the bubbles become elongated
vertically, as shown in the 2D cuts on the right of density (above)
and temperature (below).  Hot matter is effectively transported
outwards from the Galactic plane (as shown by the velocity vectors in
the top of each pair of panels on the right; the bottom of each pair
includes magnetic field vectors.
\label{korpi}}
\end{figure}

An exciting development in the last year has been the attempt by
Korpi et al.\ (1998), described by them as well in these proceedings, to
include magnetic fields in a fully three-dimensional large-scale model
of the interstellar medium including energy input from correlated and
random supernova explosions as well as the effects of galactic shear
and stratification, as shown in Figure~\ref{korpi}.

\section{Magnetic Coronas and Galactic Winds}

In addition to the interactions of magnetic fields with superbubbles
already described above, other magnetic effects on the structure of
the halo have also been proposed.  By analogy with the solar corona,
local reconnection of turbulent magnetic fields has been suggested as
an extra heating mechanism in the halo in order to explain the
anomalous line ratioes discussed above (Birk, Lesch, \& Neukirch
1998).  However this mechanism would require reconnection sheets 300
to $3 \times 10^4$ km wide throughout the halo.  Producing so much
fine structure with the relatively large-scale driving mechanisms
available such as supernova remnants and superbubbles seems difficult.
If it were indeed present, however, one might expect it to be visible
in measurements sensitive to such small scales, such as pulsar
scintillation observations.

At the other extreme of size scale, open field lines extending from
the galactic disk, produced by superbubbles blowing out, Parker
instabilities, or some combination thereof, might allow galactic winds
driven by large-scale gradients of cosmic rays (Breitschwerdt,
McKenzie, \& V\"olk 1991, 1993; Zirakashvili {\em et al.} 1996,
Ptuskin {\em et al.} 1997).  These models seem directly relevant to
the radio synchrotron haloes observed around star-forming galaxies
(e.\ g.\ Dahlem, Lisenfeld, \& Golla 1995, and see the review by
Dettmar 1992), but no direct comparisons have yet been made, to my
knowledge.  Such comparisons would be fruitful.

\section{Conclusions}


\begin{itemize}
\item Diffuse ionized gas seen in H$\alpha$ emission is most likely
ionized by dilute radiation from hot stars, perhaps with some
secondary heating from shocks or other sources.
\item Full understanding of the diffuse gas will requre models
combining the effects of dynamics and ionizing radiation.
\item Interactions between superbubbles may be as important as the
development of isolated superbubbles.
\item The Reynolds layer is probably not continuous, but rather
fragmentary and incomplete, only becoming a full layer in very active
star-forming galaxies like the Milky Way or NGC~891.
\item Superbubble blowouts transfer energy and metals to the halo and
intergalactic medium, but not significant mass.
\end{itemize}
\acknowledgments I thank the conference organizers for their
invitation and support of my attendence.  I have used the NASA
Astrophysical Data System Abstract Service in the preparation of this
review. I thank M. Avillez and M. Korpi for their provision
of figures for this review in advance of publication.

\end{document}